\documentclass[12pt,a4paper]{article}

\usepackage{iftex}
\ifPDFTeX
  \usepackage{cmap}
\fi
\usepackage[T1]{fontenc}
\usepackage{lmodern}
\usepackage{amsmath,amssymb}
\usepackage{cite}
\usepackage{graphicx}
\usepackage{placeins}
\graphicspath{{figures_fixedM/}{figures/}}
\usepackage{caption}
\usepackage[a4paper,margin=1in]{geometry}
\usepackage[colorlinks=true,linkcolor=blue,citecolor=blue,urlcolor=blue]{hyperref}

\title{Absorption and scattering of massless scalar waves by black holes with scalar hair}
\author{%
\small Lina Zhang$^{1}$\thanks{E-mail: \href{mailto:linazhang@hnit.edu.cn}{linazhang@hnit.edu.cn}},
Yun Soo Myung$^{2}$\thanks{E-mail: \href{mailto:ysmyung@inje.ac.kr}{ysmyung@inje.ac.kr} (corresponding author)},
Yu-Xuan Wang$^{3}$\thanks{E-mail: \href{mailto:wangyuxuan@jxnu.edu.cn}{wangyuxuan@jxnu.edu.cn}},
Xufen Zhang$^{4}$\thanks{E-mail: \href{mailto:xufenzhang0925@126.com}{xufenzhang0925@126.com}},\\
\small Ming Zhang$^{5}$\thanks{E-mail: \href{mailto:mingzhang0807@126.com}{mingzhang0807@126.com}},
De-Cheng Zou$^{3}$\thanks{E-mail: \href{mailto:dczou@jxnu.edu.cn}{dczou@jxnu.edu.cn}}\\[2mm]
\small $^{1}$College of Science, Hunan Institute of Technology, Hengyang 421002, China\\
\small $^{2}$Center for Quantum Spacetime, Sogang University, Seoul 04107, Republic of Korea\\
\small $^{3}$School of Physics, Jiangxi Normal University, Nanchang 330022, China\\
\small $^{4}$Department of Physics, Key Laboratory of Low Dimensional Quantum Structures\\
\small and Quantum Control of Ministry of Education, Institute of Interdisciplinary Studies,\\
\small and Synergetic Innovation Center for Quantum Effects and Applications,\\
\small Hunan Normal University, Changsha, Hunan 410081, China \\
\small $^{5}$Faculty of Science, Xi'an Aeronautical University, Xi'an 710077, China}
\date{}

\begin{document}
\maketitle

\begin{abstract}
We use partial waves to study massless scalar absorption and scattering by the normal ($\xi=5$) and phantom ($\xi=-2$) Bakopoulos--Nakas black holes, described by mass $M$ and scalar charge $q$. At fixed mass $M=1$, increasing $q/M$ decreases the horizon area and geometric capture cross section in the normal branch, but it increases both in the phantom branch.  The low-frequency absorption tends to the horizon area, while photon-sphere scaling brings the absorption spectra close together. In scattering picture, charge and branch change the interference pattern, and higher frequencies make the fringes narrow. However, matching the capture or glory scale of the normal branch with Reissner--Nordstr\"om black holes indicates  finite-frequency differences.

\end{abstract}

\section{Introduction}

Absorption and scattering of probe waves are essential for understanding the nature of black holes. The horizon boundary condition and exterior potential determine the wave response. At low frequency, the absorption cross section of a minimally coupled massless scalar approaches the horizon area~\cite{DasGibbonsMathur1997,Higuchi2001}, while at high frequency the photon sphere determines the geometric capture cross section and its leading oscillatory correction~\cite{DecaniniEspositoFareseFolacci2011}. In addition, the unstable circular null orbit sets the geodesic stability scale~\cite{CardosoEtAl2009}. These limits provide useful checks on numerical wave computations.

It is well-known that the partial wave method relates absorption to transmission probabilities and elastic scattering to phase shifts. Early calculations were carried out for Schwarzschild black holes~\cite{SanchezWave1977,SanchezAbsorption1978,SanchezScattering1978}, and a general treatment is given in Ref.~\cite{FuttermanHandlerMatzner1988}. Its applications to Reissner--Nordstr\"om (RN), Bardeen, and Hayward black holes have been described in ~\cite{CrispinoDolanOliveira2009,MacedoCrispino2014,MacedoOliveiraCrispino2015,dePaulaHayward2023}. Charged dilatonic black holes have also been studied for scalar-wave scattering~\cite{HuangZhang2020} and for both absorption and scattering in Einstein--Maxwell--dilaton theory~\cite{RicharteMartinsFabris2022}. Comparison with Schwarzschild and RN black holes have also been used to study scalar scattering by considering tidal (negative) charge~\cite{deOliveira2018}. The deflection function and backward glory approximation connect these wave calculations to  null geodesics computation~\cite{Newton1982,Matzner1985}.

On the other hand, regular black holes formed  a distinct class in absorption and scattering studies. The Bardeen and Hayward models provided  nonsingular black-hole solutions~\cite{Bardeen1968,Hayward2006}, and later works surveyed and extended this class~\cite{Ansoldi2008,Frolov2016,LanYangGuoMiao2023}. By contrast, the Bakopoulos--Nakas (BN) spacetime describes asymptotically flat black holes with scalar hair in Einstein gravity with a minimally coupled scalar and a reconstructed self-interaction potential. The original work has focused on static solutions, slow rotation, and axial perturbations~\cite{BakopoulosNakas2021}, while  exact black hole  solutions were found  with normal and phantom scalar fields~\cite{BakopoulosNakas2023}. The  latter  is considered as an very intereting black hole solution with scalar charge $q$.  In this work, we investigate  absorption and scattering in these two branches, which both may  reduce to Schwarzschild solution when the scalar charge vanishes.

For Frolov black holes, matching the critical or glory impact parameter with  RN or Hayward black hole showed  similar absorption and scattering patterns at intermediate and high frequencies~\cite{TangHuangZhang2026}.  Whether such a matching also captures the finite-frequency BN wave response is a main concern of the present work.  We therefore wish to  compare backgrounds at the same ADM mass, varying $q/M$ and determining the corresponding horizon radius. Null geodesics and scalar partial waves are then used to compare the normal and phantom branches as well as  to compare normal BN and RN black holes at independently matched capture and glory scales.

Our work is organized as follows. Section~\ref{sec:geometry} is devoted to performing  the geometry and classical analysis. In Section~\ref{sec:partial-wave}, we develop the partial  wave calculation. Numerical results and comparisons appear in Section~\ref{sec:results}. Finally, we conclude in Section~\ref{sec:conclusion}.

\section{The Bakopoulos-Nakas (BN) geometry and classical analysis}
\label{sec:geometry}

\subsection{BN black hole and null geodesics}
\begin{figure}[htbp]
  \centering
  \includegraphics[width=0.70\linewidth]{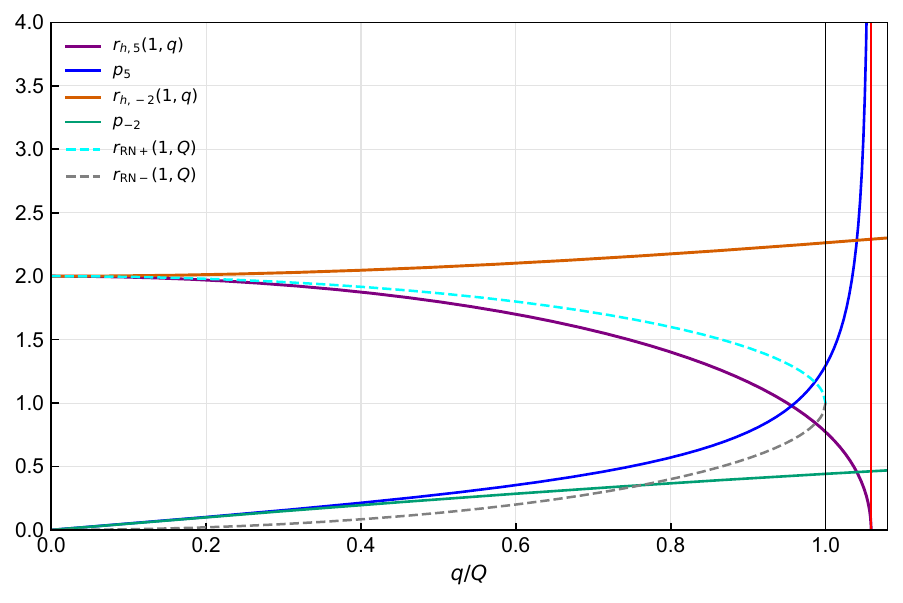}
  \caption{BN horizon radii $r_{h,5}(1,q)$ and $r_{h,-2}(1,q)$, ratios $p_\xi=q/r_{h,\xi}$, and RN horizons at $M=1$. The RN curves are $r_{\mathrm{RN}\pm}(1,Q)=1\pm\sqrt{1-Q^2}$, with plus and minus denoting the outer and inner horizons. The horizontal label $q/Q$ denotes $q$ for BN and $Q$ for RN. The normal BN branch exists for $0\leq q<18/17$, with a vanishing horizon at the red limiting line, whereas the phantom horizon grows with $q$ without limitation. The black line marks RN extremality at $Q=1$, where both horizons equal unity. $p_\xi$ curves denote charge-to-horizon ratios.}
  \label{fig:horizon-r}
\end{figure}

Let us introduce the Einstein-scalar action with minimal coupling~\cite{BakopoulosNakas2021}
\begin{equation}
    S_\xi=\frac{1}{16\pi}\int d^4x\sqrt{-g}\Big[R-\frac{\varepsilon_\xi}{2}(\partial \Phi_\xi)^2-V_\xi(\Phi_\xi)\Big],
    \qquad \varepsilon_5=+1,\quad \varepsilon_{-2}=-1
\end{equation}
Here, the scalar potential $V_\xi(\Phi_\xi)$ was reconstructed together with each BN solution. The sign $\varepsilon_\xi=-1$ gives the phantom kinetic term and  in this case, the field in the action becomes the real field $\Phi_{-2}\equiv\widetilde\Phi_{-2}$ defined below.
For a static and spherically symmetric configuration, we introduce the BN line element as
\begin{equation}
 \mathrm{d}s^2_{\rm BN}=g_{\mu\nu}dx^\mu dx^\nu=-F(r)\,\mathrm{d}t^2+\frac{\mathrm{d}r^2}{B(r)}
 +r^2\left(\mathrm{d}\theta^2+\sin^2\theta\,\mathrm{d}\varphi^2\right),
 \qquad F(r)=e^{A(r)}B(r).
 \label{eq:metric-general}
\end{equation}
The normal scalar profile, the real phantom profile, and the corresponding functions $A_\xi(r)$ are given by
\begin{equation}
 \begin{aligned}
 \Phi_5(r)&=2\sqrt{5}\,\operatorname{arsinh}\!\left(\frac{q}{r}\right),&
 \widetilde\Phi_{-2}(r)&=2\sqrt{2}\,\operatorname{arsinh}\!\left(\frac{q}{r}\right),\\
 A_5(r)&=-5\ln s,& A_{-2}(r)&=2\ln s,\qquad s=1+\frac{q^2}{r^2}.
 \end{aligned}
 \label{eq:scalar-profile}
\end{equation}
We note that for $\xi=-2$, the original scalar profile is imaginary but the real field $\widetilde\Phi_{-2}$  may be allowed  after making a redefinition of the phantom field~\cite{BakopoulosNakas2023}.  For a compact representation, let us define
\begin{equation}
 \begin{aligned}
 z&=\frac{q^2}{r^2},\qquad s=1+z,\\
 D_5(r)&=1+3z+\frac{17}{9}z^2-\frac{2M}{r}s^{3/2},\\
 D_{-2}(r)&=1-\frac{2M}{r}\left(1+\frac{3z}{5}\right)-\frac{z^2}{3}.
 \end{aligned}
 \label{eq:zsD}
\end{equation}
The two exact branches have the different metric functions as 
\begin{equation}
 F_5=\frac{D_5}{s^3},\quad B_5=s^2D_5,\qquad
 F_{-2}=D_{-2},\quad B_{-2}=\frac{D_{-2}}{s^2}.
 \label{eq:BN-functions}
\end{equation}
Here $M$ is the ADM mass and $q$ is the scalar-hair charge, being distinct from an electromagnetic charge.  The normal branch has a canonical kinetic term, while the phantom branch has the opposite sign. At $q=0$, both metrics recover Schwarzschild solution. We use $G=c=\hbar=1$ and signature $(-,+,+,+)$. The event horizon $r_h(M,q)$ is the positive root of the corresponding $D_\xi(r_h)=0$. Introducing $p=q/r_h$ gives a convenient parametric representation:

\begin{equation}
 \begin{aligned}
 \frac{r_h}{M}&=\frac{2(1+p^2)^{3/2}}{1+3p^2+\frac{17}{9}p^4},&
 \frac{q}{M}&=\frac{2p(1+p^2)^{3/2}}{1+3p^2+\frac{17}{9}p^4},&& (\xi=5),\\
 \frac{r_h}{M}&=\frac{2(1+3p^2/5)}{1-p^4/3},&
 \frac{q}{M}&=\frac{2p(1+3p^2/5)}{1-p^4/3},&& (\xi=-2),\quad 0\le p<3^{1/4}.
 \end{aligned}
 \label{eq:horizon-param}
\end{equation}
The BN horizon radii and charge-to-horizon ratios are shown alongside the RN horizons in Fig.~\ref{fig:horizon-r}. For equal  charges in the common plotted range, the normal BN horizon ($0\le r_{h,5}\le 2$) lies below the RN outer horizon, while the phantom horizon ($r_{h,-2}\ge 2$) lies above it. This feature will show distinct null geodesic pictures for two branches. The RN horizons approach each other as $Q$ increases and meet at extremality. Both BN charge-to-horizon ratios grow with $q$, more rapidly on the normal branch. Especially, we note that there is no upper bound on $q$ for the phantom BN horizon radius. 
All comparisons fix the ADM mass at $M=1$, with $r_h/M$ determined by $D_\xi(r_h)=0$ for each branch and charge. We use $M$ as the common length scale, so the displayed hair charge  and frequency parameters will be given by $q/M$ and $\omega M$. Varying $q$ also changes the reconstructed background scalar potential, so this is not a charge variation within one independently fixed scalar self-interaction theory.

Because the metric functions depend on $q^2$, changing the sign of $q$ reverses the scalar profile but leaves the probe observables unchanged, so choosing $q\geq0$ is sufficient. On the normal branch, $q/M$ approaches $18/17$ as $p\to\infty$ while $r_h/M\to0$. The phantom branch instead has $0\leq p<3^{1/4}$ and no upper bound on $q/M$. The sampled horizons are nonextremal such that for the normal branch $D'_{5,h}=(1+p^2+p^4/3)/[r_h(1+p^2)]>0$, while for the phantom branch $D'_{-2,h}=2M r_h^{-2}(1+9p^2/5)+4p^4/(3r_h)>0$.

To study equatorial null geodesics, the conserved energy $E$, angular momentum $L$, and impact parameter $b$ are defined as 
\begin{equation}
 E=F(r)\dot{t},\qquad L=r^2\dot{\varphi},\qquad b=\frac{L}{E},
 \label{eq:EL}
\end{equation}
where a dot denotes differentiation with respect to affine parameter $\tau$. The radial equation leads to
\begin{equation}
 \dot{r}^{\,2}=B(r)\left[\frac{E^2}{F(r)}-\frac{L^2}{r^2}\right].
 \label{eq:null-radial}
\end{equation}
The unstable circular null orbit $r_c$, critical impact parameter $b_c$, and geometric capture cross section $\sigma_{\rm geo}$ are given by
\begin{equation}
 r_cF'(r_c)-2F(r_c)=0,\qquad
 b_c=\frac{r_c}{\sqrt{F(r_c)}},\qquad
 \sigma_{\rm geo}=\pi b_c^2.
\label{eq:photon-sphere}
\end{equation}
The angular frequency $\Omega_c$ and Lyapunov exponent $\lambda_c$ defined with respect to the coordinate time $t$ take the forms
\begin{equation}
 \Omega_c=\frac{\sqrt{F_c}}{r_c}=\frac{1}{b_c},\qquad
 \lambda_c=\left[\frac{B_c(2F_c-r_c^2F_c'')}{2r_c^2}\right]^{1/2},
 \label{eq:Omega-lambda}
\end{equation}
where a subscript $c$ denotes evaluation at $r=r_c$. These determine the capture scale and enter the high-frequency absorption approximation. For both exact BN branches, evaluating these expressions at the circular null orbit gives $\lambda_c/\Omega_c=1$ throughout the sampled family.

The optical potential $U_\xi=F_\xi/r^2$ determines capture through $b_{c,\xi}^{-2}=\max_{r>r_{h,\xi}}U_\xi$. With $U_{\rm S}=(1-2M/r)/r^2$ and $\rho=\sqrt{r^2+q^2}$, Eq.~\eqref{eq:BN-functions} gives
\begin{equation}
 \begin{aligned}
 U_{-2}&=U_{\rm S}-\frac{6Mq^2}{5r^5}-\frac{q^4}{3r^6}
 \leq\frac{1}{27M^2},\\
 \left.U_5\right|_{\rho=3M}&=\frac{1}{27M^2}+\frac{q^2}{81M^4}
 \left(1-\frac{q^2}{81M^2}\right)\geq\frac{1}{27M^2}.
 \end{aligned}
 \label{eq:branch-optical-bounds}
\end{equation}
The point $\rho=3M$ is located outside the normal horizon since $r_{h,5}\leq2M$ and $q/M<18/17$. Hence, we have 
\begin{equation}
 \frac{b_{c,5}}{M}\leq3\sqrt{3}\simeq5.1962,\qquad
 \frac{b_{c,-2}}{M}\geq3\sqrt{3}
 \label{eq:branch-capture-bounds}
\end{equation}
with equality holding only at $q=0$, the common Schwarzschild limit.
\begin{figure}[!htbp]
  \centering
  \includegraphics[width=0.98\linewidth]{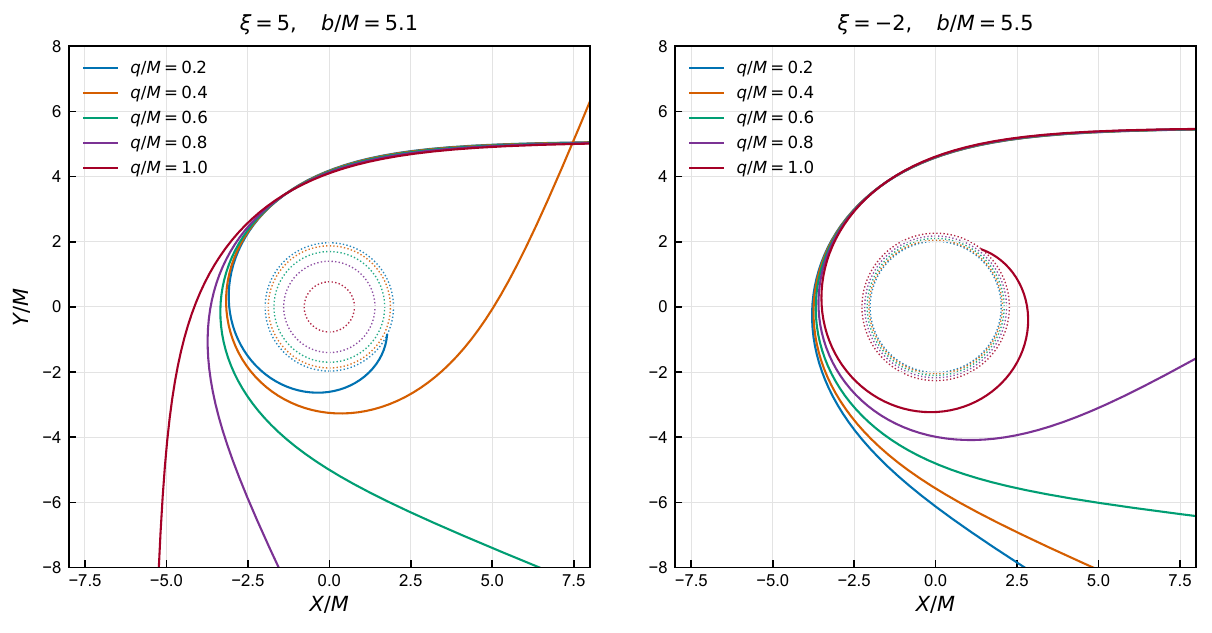}
  \caption{Null geodesics in $(X/M,Y/M)$, with $X=r\cos\varphi$ and $Y=r\sin\varphi$, at $M=1$. The left and right panels show $\xi=5$ and $\xi=-2$, respectively. Both panels use $q/M=0.2,0.4,0.6,0.8,1.0$, with $b/M=5.1$ for $\xi=5$ and $5.5$ for $\xi=-2$. Only the $q/M=0.2$ and $1.0$ rays, respectively, are captured. Colored dotted circles mark the charge-dependent horizons. The different fixed $b/M$ values are required by the different critical capture scales of the two branches.}
  \label{fig:null-geodesics}
\end{figure}
At fixed $M=1$, capture requires $b<b_c$. A common $b/M$ that captures any normal-branch ray therefore captures all phantom-branch rays. Conversely, if any phantom-branch ray escapes, all normal-branch rays escape. Thus, Fig.~\ref{fig:null-geodesics} uses $b/M=5.1$ for $\xi=5$ and $5.5$ for $\xi=-2$, giving one captured ray in each panel. With the same uniformly spaced charge samples, increasing $q/M$ takes the normal-branch rays from capture to escape, whereas the phantom-branch rays pass from escape to capture. These opposite trends follow from the decrease and increase of $b_c$ in the two branches. This shows clearly two distinct pictures for normal and phantom branches.

\subsection{Weak deflection and classical scattering picture}

For a scattering ray with $b>b_c$, the turning point $r_0$ is determined by the relation of $b^2F(r_0)=r_0^2$. The deflection angle is given by
\begin{equation}
\Theta(b)=2\int_{r_0}^{\infty}\frac{\mathrm{d}r}{r^2\sqrt{B(r)}}
\left[\frac{1}{b^2F(r)}-\frac{1}{r^2}\right]^{-1/2}-\pi.
\label{eq:deflection}
\end{equation}
On the principal nonwinding branch, the scattering angle is $\vartheta=\Theta(b)\in(0,\pi)$, being distinct from the spherical-polar coordinate $\theta$ in Eq.~\eqref{eq:metric-general}. For winding rays, $\Theta$ needs not equal the observed angle.  Actually, these additional branches are not included in the classical comparison below. We note that angles in the analytic formulas are in radians, whereas the plotted angular axes use degrees.
For the normal ($\xi=5$) branch, the asymptotic metric functions take the forms
\begin{equation}
 F(r)\simeq 1-\frac{2M}{r}+\frac{3Mq^2}{r^3}+O\Big(\frac{1}{r^{4}}\Big),\qquad
 B(r)\simeq1-\frac{2M}{r}+\frac{5q^2}{r^2}-\frac{7Mq^2}{r^3}+O\big(\frac{1}{r^{4}}\Big).
 \label{eq:asymptotic-metric}
\end{equation}
Expanding the normal-branch orbit integral with these coefficients gives
\begin{equation}
 \Theta(b)\simeq \frac{4M}{b}
 +\frac{\pi(15M^2-5q^2)}{4b^2}
 +O\!\left(\frac{M^3+Mq^2}{b^3},\frac{q^4}{b^4}\right).
 \label{eq:weak-deflection}
\end{equation}
For this branch, the scalar-hair contribution first enters at order $1/b^{2}$, with a coefficient differing from the RN charge term. Since the expansion assumes $b\gg M,|q|$, it describes the weak-deflection branch. The classical differential cross section follows from the deflection function as
\begin{equation}
 \frac{\mathrm{d}\sigma_{\rm cl}}{\mathrm{d}\Omega}
 =\frac{b}{\sin\vartheta}\left|\frac{\mathrm{d}b}{\mathrm{d}\vartheta}\right|,
 \label{eq:classical-dcs}
\end{equation}
where $\vartheta$ denotes the scattering angle~\cite{Newton1982}. In the small angle limit, inversion of Eq.~\eqref{eq:weak-deflection} gives
\begin{equation}
 \frac{\mathrm{d}\sigma_{\rm cl}}{\mathrm{d}\Omega}
 =\frac{16M^2}{\vartheta^4}
 +\frac{\pi(15M^2-5q^2)}{4\vartheta^3}
 +O\Big(\frac{1}{\vartheta^{2}}\Big),
 \label{eq:small-angle-dcs}
\end{equation}
where $\vartheta$ is measured in radians. For the phantom branch of $F_{-2}=1-2M/r+O(r^{-3})$ and $B_{-2}=1-2M/r-2q^2/r^2+O(r^{-3})$,  the asymptotic coefficients instead give
\begin{equation}
 \begin{aligned}
 \Theta_{-2}(b)&=\frac{4M}{b}+\frac{\pi(15M^2+2q^2)}{4b^2}
 +O\!\left(\frac{M^3+Mq^2}{b^3},\frac{q^4}{b^4}\right),\\
 \frac{\mathrm{d}\sigma_{{\rm cl},-2}}{\mathrm{d}\Omega}
 &=\frac{16M^2}{\vartheta^4}+\frac{\pi(15M^2+2q^2)}{4\vartheta^3}+O(\vartheta^{-2}).
 \end{aligned}
 \label{eq:phantom-weak-deflection}
\end{equation}
Thus, the first hair-dependent correction shows opposite signs in the two branches. The classical curves for both branches are calculated from their full deflection functions.

Near backward scattering, the leading glory approximation (interference approximation associated with trajectories deflected by $\pi$)  is expressed as 
\begin{equation}
 \left.\frac{\mathrm{d}\sigma_{\rm sc}}{\mathrm{d}\Omega}\right|_{\rm glory}
 \simeq 2\pi\omega b_g^2
 \left|\frac{\mathrm{d}b}{\mathrm{d}\vartheta}\right|_{\vartheta=\pi}
 J_0^2\!\left(\omega b_g\sin\vartheta\right),
 \label{eq:glory}
\end{equation}
where $b_g$ is the impact parameter for which the deflection angle is $\pi$ and $J_0$ is the Bessel function of the first kind~\cite{Matzner1985}.

\section{Partial-wave approach}
\label{sec:partial-wave}

\subsection{Massless scalar propagation}

The wave field $\Psi$ is considered as  an independent, minimally coupled massless probe, while the background profile $\Phi$ is held fixed.  It is not the  scalar perturbation $\delta\Phi$, which would couple to the metric and background self-interaction. The probe obeys the Klein--Gordon equation 
\begin{equation}
 \bar{\Box}\Psi=\frac{1}{\sqrt{-g}}\partial_\mu
 \left(\sqrt{-g}\,g^{\mu\nu}\partial_\nu\Psi\right)=0.
 \label{eq:KG}
\end{equation}
Separating variables and using the spherical-harmonic eigenvalue equation gives 
\begin{equation}
 \Psi(t,r,\theta,\varphi)=\frac{u_\ell(r)}{r}
 Y_{\ell m}(\theta,\varphi)e^{-i\omega t},\qquad
 \nabla^2_{S^2}Y_{\ell m}=-\ell(\ell+1)Y_{\ell m}
\label{eq:separation}
\end{equation}
together with the tortoise coordinate
\begin{equation}
 \frac{\mathrm{d}r_*}{\mathrm{d}r}=\frac{1}{\sqrt{F(r)B(r)}}.
 \label{eq:tortoise-general}
\end{equation}
Then, the radial function satisfies a Schr\"odinger-type equation
\begin{equation}
 \frac{\mathrm{d}^2u_\ell}{\mathrm{d}r_*^2}
 +\left[\omega^2-V_\ell(r)\right]u_\ell=0,
 \qquad
 V_\ell(r)=F(r)\frac{\ell(\ell+1)}{r^2}
 +\frac{[F(r)B(r)]'}{2r}.
 \label{eq:radial-master}
\end{equation}
Since spherical symmetry makes the transmission probability independent of $m$, the factor $(2\ell+1)$ in the cross sections represents the $m$-degeneracy of the incident unit-amplitude plane wave.
For the normal branch with $F_5B_5=D_5^2/s$, so the tortoise coordinate and $D'_5(r)$ are
\begin{equation}
 \frac{\mathrm{d}r_*}{\mathrm{d}r}=\frac{\sqrt{s}}{D_5},\qquad
 D'_5(r)=-\frac{6z}{r}-\frac{68z^2}{9r}
 +\frac{2M}{r^2}\sqrt{s}(1+4z).
 \label{eq:tortoise-xi5}
\end{equation}
The normal-branch effective potential is expressed compactly as 
\begin{equation}
 V_{\ell,5}(r)=\frac{D_5\ell(\ell+1)}{r^2s^3}
 +\frac{D_5D'_5}{rs}+\frac{zD_5^2}{r^2s^2}.
 \label{eq:potential-xi5}
\end{equation}
For the phantom branch with $F_{-2}B_{-2}=D_{-2}^2/s^2$, one has 
\begin{equation}
 \begin{aligned}
 \frac{\mathrm{d}r_*}{\mathrm{d}r}&=\frac{s}{D_{-2}},\qquad
 D'_{-2}(r)=\frac{2M}{r^2}\left(1+\frac{9z}{5}\right)+\frac{4z^2}{3r},\\
 V_{\ell,-2}&=\frac{D_{-2}\ell(\ell+1)}{r^2}
 +\frac{D_{-2}D'_{-2}}{rs^2}+\frac{2zD_{-2}^2}{r^2s^3}.
 \end{aligned}
 \label{eq:potential-phantom}
\end{equation}
The first term in each potential is the angular momentum barrier, while the remaining terms contribute to the $\ell=0$ mode. Increasing $\ell$ raises the barrier at fixed background. The potentials are non-negative outside the horizon in the sampled cases, consistent with the absence of growing modes in the minimally coupled probe-scalar theory.  However, this result alone does not establish dynamic stability against coupled perturbations of the metric and the background scalar field.

\begin{figure}[htbp]
  \centering
  \includegraphics[width=0.96\linewidth]{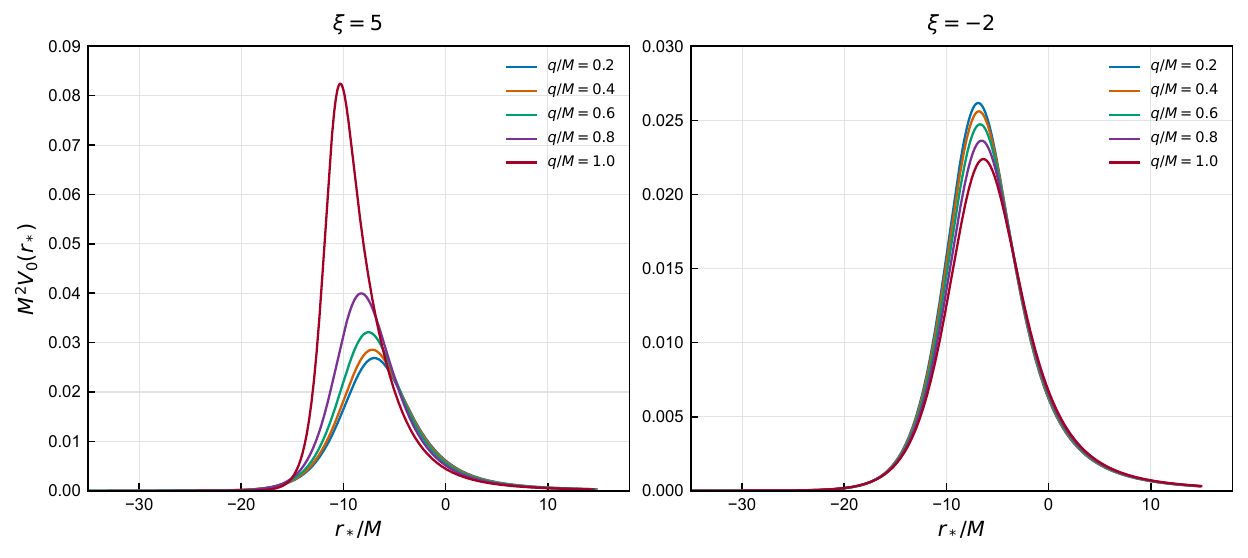}
  \caption{Effective potential $M^2V_0(r_*)$ for $\ell=0$ and $M=1$. Left: $\xi=5$, while  right: $\xi=-2$. Both panels use $q/M=0.2,0.4,0.6,0.8,1.0$, with the same charge colors. The horizontal scale is shared, but the vertical ranges differ. Each curve uses $r_*(6M)=0$, aligning the tortoise origin at a common physical radius.}
  \label{fig:potential-coordinates}
\end{figure}

Fig.~\ref{fig:potential-coordinates} displays  the $\ell=0$ barriers in separate panels at $M=1$ with the same charge samples in both branches. The normal-branch peak rises with increasing $q/M$, while the phantom-branch peak falls.

\subsection{Boundary conditions and probabilities}
The boundary conditions are purely ingoing at the horizon and a superposition of incoming and outgoing waves at infinity. Near the horizon, $D_\xi(r_h)=0$ gives the tortoise coordinate and surface gravity
\begin{equation}
 r_*\simeq C_\xi\ln\!\left(\frac{r-r_h}{r_h}\right),\qquad
 \kappa_\xi=\frac{1}{2C_\xi},\qquad
 C_5=\frac{\sqrt{s_h}}{D'_{5,h}},\quad C_{-2}=\frac{s_h}{D'_{-2,h}}.
 \label{eq:horizon-rstar}
\end{equation}
The purely ingoing solution is given by 
\begin{equation}
 u_\ell\sim A^{\rm tr}_\ell e^{-i\omega r_*},\qquad r\to r_h^+.
 \label{eq:horizon-bc}
\end{equation}
At infinity, the asymptotically flat metric gives
\begin{equation}
 r_*\simeq r+2M\ln(r/M)+C_*+O(r^{-1}),\qquad
 u_\ell\sim A^{\rm in}_\ell e^{-i\omega r_*}
 +A^{\rm out}_\ell e^{+i\omega r_*}.
\label{eq:infinity-bc}
\end{equation}
The additive constant $C_*$ specifies the chosen origin of the tortoise coordinate.
The partial-wave $S$ matrix and transmission probability $\Gamma_\ell$ are defined as 
\begin{equation}
 S_\ell=(-1)^{\ell+1}\frac{A^{\rm out}_\ell}{A^{\rm in}_\ell},\qquad
 \Gamma_\ell=1-|S_\ell|^2
 =\frac{|A^{\rm tr}_\ell|^2}{|A^{\rm in}_\ell|^2}.
 \label{eq:S-Gamma}
\end{equation}

\subsection{Absorption cross sections}

The total absorption cross section is defined  by  summing  over partial waves as 
\begin{equation}
 \sigma_{\rm abs}(\omega)=\sum_{\ell=0}^{\infty}\sigma_\ell(\omega),\qquad
 \sigma_\ell(\omega)=\frac{\pi}{\omega^2}(2\ell+1)\Gamma_\ell(\omega).
 \label{eq:absorption}
\end{equation}
Absorption cross sections are displayed in units of $\pi M^2$, with frequency measured by $\omega M$.
The low-frequency limit equals the horizon area $\mathcal A_h$ as 
\begin{equation}
 \lim_{\omega M\to0}\sigma_{\rm abs}=\mathcal A_h=4\pi r_h^2.
 \label{eq:low-frequency}
\end{equation}
At high frequency, the leading term denotes the geometric capture cross section. The photon-sphere correction takes the form~\cite{DecaniniEspositoFareseFolacci2011}
\begin{equation}
 \sigma_{\rm hf}(\omega)\simeq\sigma_{\rm geo}
 \left[1-8\pi\beta_c e^{-\pi\beta_c}
 \operatorname{sinc}(2\pi b_c\omega)\right],\qquad
 \beta_c=\frac{\lambda_c}{\Omega_c}
 \label{eq:sinc}
\end{equation}
with $\operatorname{sinc}x=\sin x/x$. This eikonal approximation requires $\omega b_c\gg1$. Where extended beyond this regime, it serves only as a reference.

\subsection{Scattering cross section}

The scalar scattering amplitude and differential cross section are given by
\begin{equation}
 f(\vartheta)=\frac{1}{2i\omega}\sum_{\ell=0}^{\infty}(2\ell+1)
 [S_\ell(\omega)-1]P_\ell(\cos\vartheta),\qquad
 \frac{\mathrm{d}\sigma_{\rm sc}}{\mathrm{d}\Omega}=|f(\vartheta)|^2.
 \label{eq:scattering}
\end{equation}
The phase shifts in $S_\ell$ encode wave effects which are absent for the classical deflection function. Near the forward direction, the $1/r$ tail produces Coulomb-like long range behavior and a finite partial-wave sum converges slowly.  The Yennie-Ravenhall-Wilson (YRW) reduction improves this numerical convergence, but it does not remove the physical forward phase~\cite{YennieRavenhallWilson1954,DolanDoranLasenby2006}. The common plotted interval is determined as  $20^\circ\leq\vartheta\leq179.8^\circ$. The lower bound avoids the long-range forward tail, while the upper  bound  excludes the $1/\sin\vartheta$ divergence of the classical reference at $\vartheta=\pi$. We note that  the scalar-wave result and glory approximation remain finite there. The classical curves use the turning-point root on the principal nonwinding branch. Captured rays with $b<b_c$ and winding branches are excluded from this classical reference.  The partial-wave result includes the full wave response.

\section{Numerical results}
\label{sec:results}

\subsection{Numerical scheme}
We briefly describe the numerical scheme.
The radial system is integrated in $r_*$, with $r$ evolved through the relation of $\mathrm{d}r=\sqrt{FB}\mathrm{d}r_*$. For absorption, integration starts at $r_{\rm start}=r_h(1+\epsilon)$ with the purely ingoing horizon expansion and $\epsilon=10^{-6}$. The default matching radius is given by
\begin{equation*}
 r_{\rm match}=\max\!\left(240r_h,\frac{1.8(\ell_{\max}+1)}{\omega},35M\right),
\end{equation*}
and we use $\mathrm{rtol}=2\times10^{-9}$, $\mathrm{atol}=2\times10^{-11}$, and $\Delta r_*\leq\min[r_h,\pi/(12\omega)]$. At $r_{\rm match}$, $u_\ell$ and $\mathrm{d}u_\ell/\mathrm{d}r_*$ are decomposed into incoming and outgoing Riccati-Hankel basis functions with the asymptotic ADM logarithmic phase. Integration and matching share the origin $r_*(r_{\rm start})=0$. The zero-frequency entries are set by the analytic horizon-area limit.

Selected transmission probabilities agree with an independent integration in horizon units after frequency rescaling. Low-frequency results approach the horizon-area limit in both branches. Absorption convergence is checked by increasing $\ell_{\max}$ from $10$ to $16$ at $\omega M=1.6$ for the two endpoint charges. Flux balance is monitored using
\begin{equation}
 \delta_\ell^{\rm flux}
 =\left|1-|S_\ell|^2-\left|
 \frac{A_\ell^{\rm tr}}{A_\ell^{\rm in}}\right|^2\right|.
 \label{eq:flux-residual}
\end{equation}

The total absorption sums use $\ell_{\max}=10$. The scattering sums use three YRW reductions and $\ell_{\max}=240$, with the additional radial modes required by the recurrence. These integrations use $\epsilon=10^{-8}$, $\mathrm{rtol}=2\times10^{-11}$, $\mathrm{atol}=2\times10^{-13}$, $\Delta r_*\leq\min[r_h,\pi/(18\omega)]$, and $r_{\rm match}=3840r_h$. The scattering convergence checks increase the cutoff from $220$ to $240$, compare three with four YRW reductions, and double the matching radius. The cross sections agree to within $0.02\%$ in these checks.

\subsection{Scalar absorption}
\begin{figure}[htbp]
  \centering
  \includegraphics[width=0.96\linewidth]{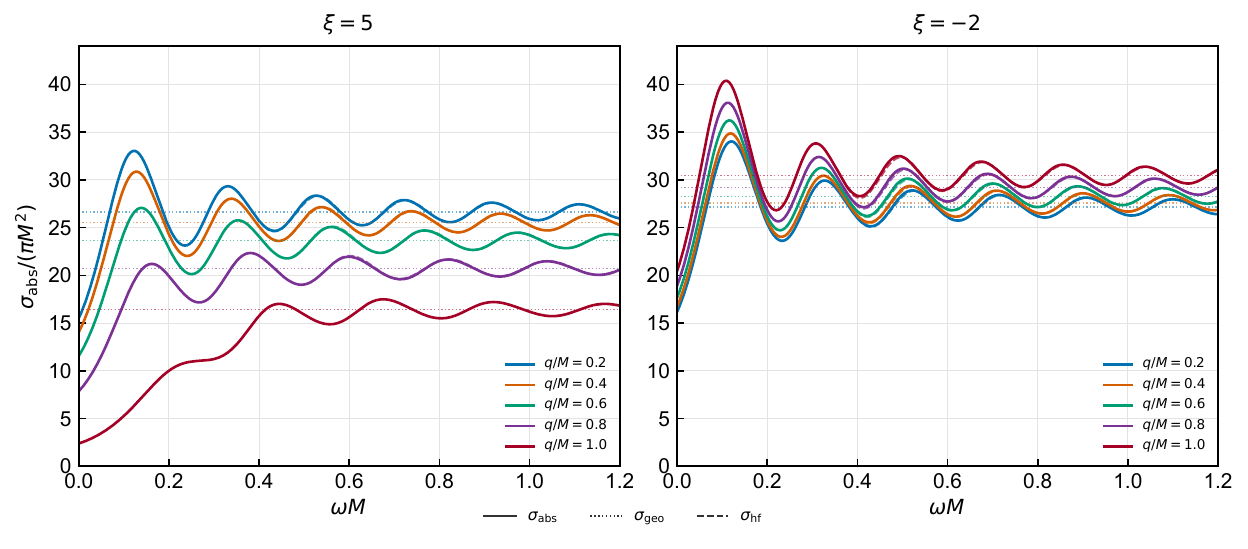}
  \caption{Total absorption $\sigma_{\rm abs}/(\pi M^2)$ as a function of $\omega M$ at $M=1$ for $q/M=0.2,0.4,0.6,0.8,1.0$. The left and right panels show $\xi=5$ and $\xi=-2$, respectively. Solid curves are numerical spectra and  dotted horizontal lines give $\sigma_{\rm geo}/(\pi M^2)$. Dashed curves show the leading high-frequency reference where $\omega r_h\geq0.8$, a charge-dependent threshold in $\omega M$. The zero-frequency entries are the analytic limits $4(r_h/M)^2$.}
  \label{fig:total-absorption}
\end{figure}

Here, we study scalar absorption by the BN black holes. At fixed mass, the normal and phantom branches respond oppositely as $q/M$ increases.  The horizon area and capture scale decrease in the normal branch, whereas both increase in the phantom branch. The total spectra in Fig.~\ref{fig:total-absorption} oscillate about the corresponding capture limits. Their zero-frequency values are fixed by the horizon areas. At higher frequencies, the reference follows the main numerical oscillations in both panels.

\begin{figure}[htbp]
  \centering
  \includegraphics[width=0.88\linewidth]{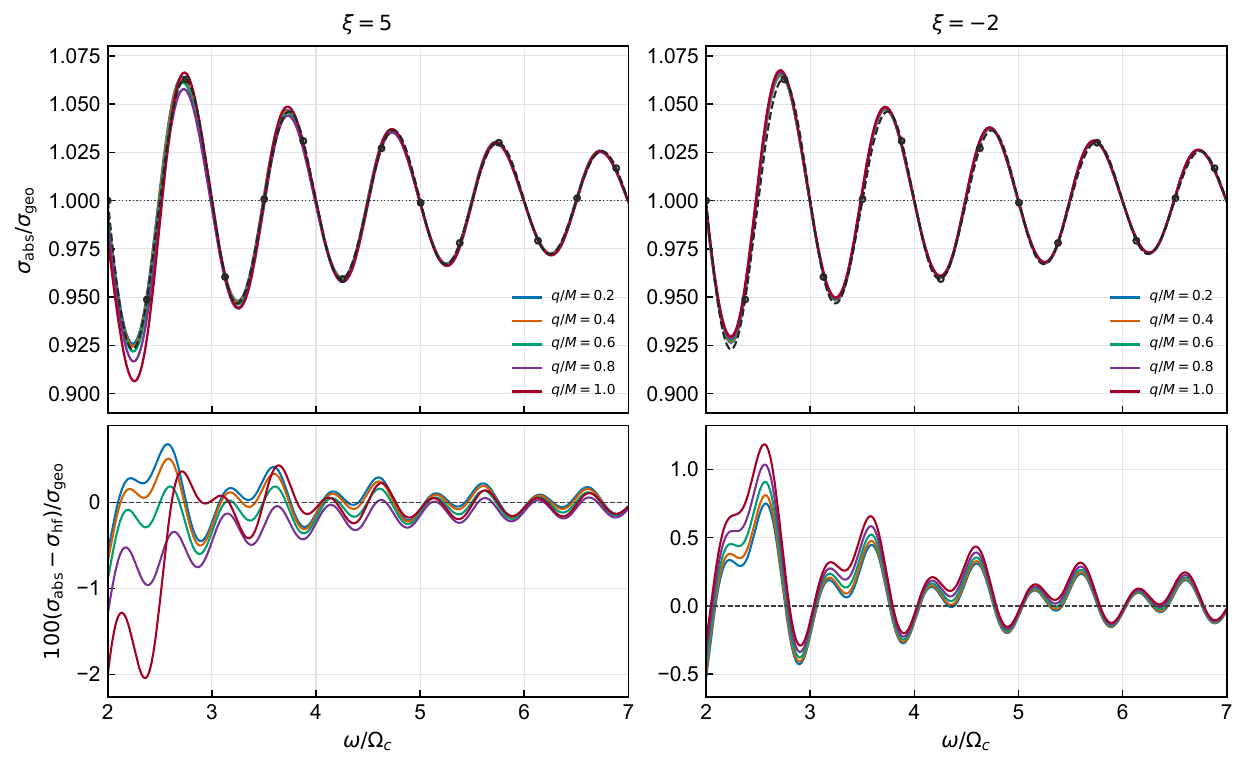}
  \caption{Photon-sphere scaled absorption for $M=1$ and $q/M=0.2,0.4,0.6,0.8,1.0$, with $\Omega_c=1/b_c$ and $2\leq\omega/\Omega_c\leq7$.  Upper: colored solid curves give $\sigma_{\rm abs}/\sigma_{\rm geo}$,  the black dashed curve with open circles denotes the common leading high-frequency reference, and the dotted line marks unity. Lower: $100(\sigma_{\rm abs}-\sigma_{\rm hf})/\sigma_{\rm geo}$, using the same charge colors.}
  \label{fig:scaled-absorption}
\end{figure}

It is important to compare the absorption spectra with the photon-sphere reference. In Fig.~\ref{fig:scaled-absorption}, plotting $\sigma_{\rm abs}/\sigma_{\rm geo}$ against $\omega/\Omega_c$ brings the spectra close together in both branches. At fixed $\omega/\Omega_c$, $\omega M$ varies with charge and branch. Since $\beta_c=\lambda_c/\Omega_c=1$, the leading high-frequency reference is common to both branches and all five charges. The lower panels resolve the remaining differences: increasing $q/M$ deepens the negative residual near the low end in the normal branch, but strengthens the first positive residual peak in the phantom branch. The residual amplitudes are smaller toward the upper end of the plotted range. The low end, $\omega/\Omega_c=2$, lies outside the strict eikonal regime.

\begin{figure}[htbp]
  \centering
  \includegraphics[width=0.96\linewidth]{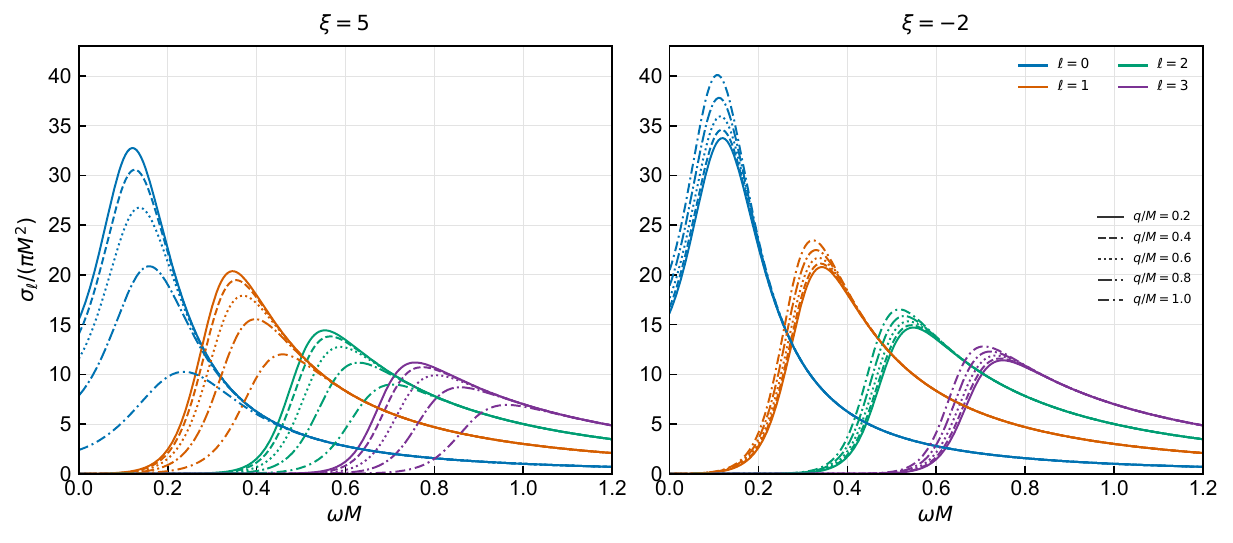}
  \caption{Partial absorption $\sigma_\ell/(\pi M^2)$ as a function of $\omega M$ at $M=1$. The left and right panels show $\xi=5$ and $\xi=-2$, respectively. Colors identify $\ell=0,1,2,3$ and  line styles distinguish $q/M=0.2,0.4,0.6,0.8,1.0$. The analytic zero-frequency values are located  $\sigma_0/(\pi M^2)=4(r_h/M)^2$ and $\sigma_{\ell>0}=0$.}
  \label{fig:partial-absorption}
\end{figure}

Finally, the individual modes in Fig.~\ref{fig:partial-absorption} explain how the absorption spectrum develops. Higher-$\ell$ modes peak at higher frequencies. At fixed $\ell$, increasing $q/M$ lowers the normal-branch peak and shifts it to higher $\omega M$, whereas the phantom-branch peak becomes higher and moves to lower $\omega M$. These trends are consistent with the opposite changes in the potential barriers. The $\ell=0$ mode supplies the horizon-area limit, while higher modes become important as the frequency increases. Once a mode becomes nearly fully transmitted, its partial cross section decreases as $1/\omega^2$.

\FloatBarrier
\subsection{Scalar scattering }
\begin{figure}[!htbp]
  \centering
  \includegraphics[width=0.98\linewidth]{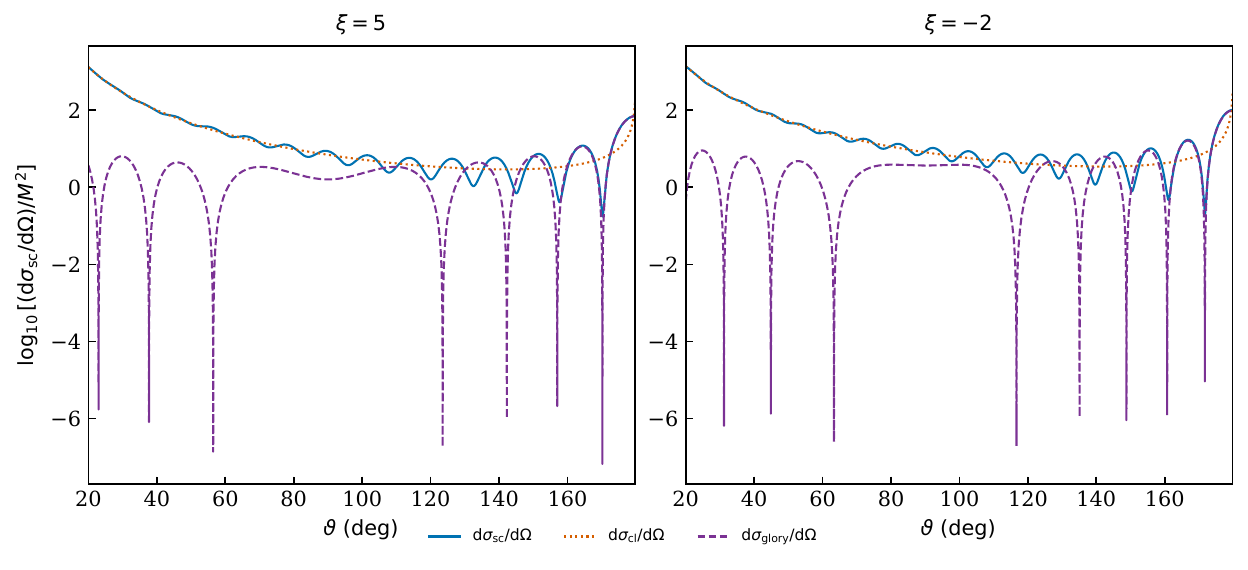}
  \captionof{figure}{Differential scattering $\log_{10}[M^{-2}\mathrm{d}\sigma_{\rm sc}/\mathrm{d}\Omega]$ as a function of $\vartheta$ at $M=1$, $q/M=0.8$, and $\omega M=3$. The left and right panels show $\xi=5$ and $\xi=-2$, respectively. The blue solid curves give the partial-wave result. Orange dotted and purple dashed curves give the principal-branch classical result and the leading backward glory approximation. The angular interval is $20^\circ\leq\vartheta\leq179.8^\circ$.}
  \label{fig:differential-scattering}
\end{figure}

In this section, we study scalar scattering by the BN black holes. In Fig.~\ref{fig:differential-scattering}, the classical reference follows the overall small-angle trend but contains no interference fringes. Near backward scattering, the glory approximation follows the fringe positions but predicts much deeper minima than the partial-wave result. At the same charge and frequency, the phantom branch has a higher backward peak and more closely spaced fringes. The extension of the glory approximation to smaller angles serves only as a reference.

\begin{figure}[!htbp]
  \centering
  \includegraphics[width=0.98\linewidth]{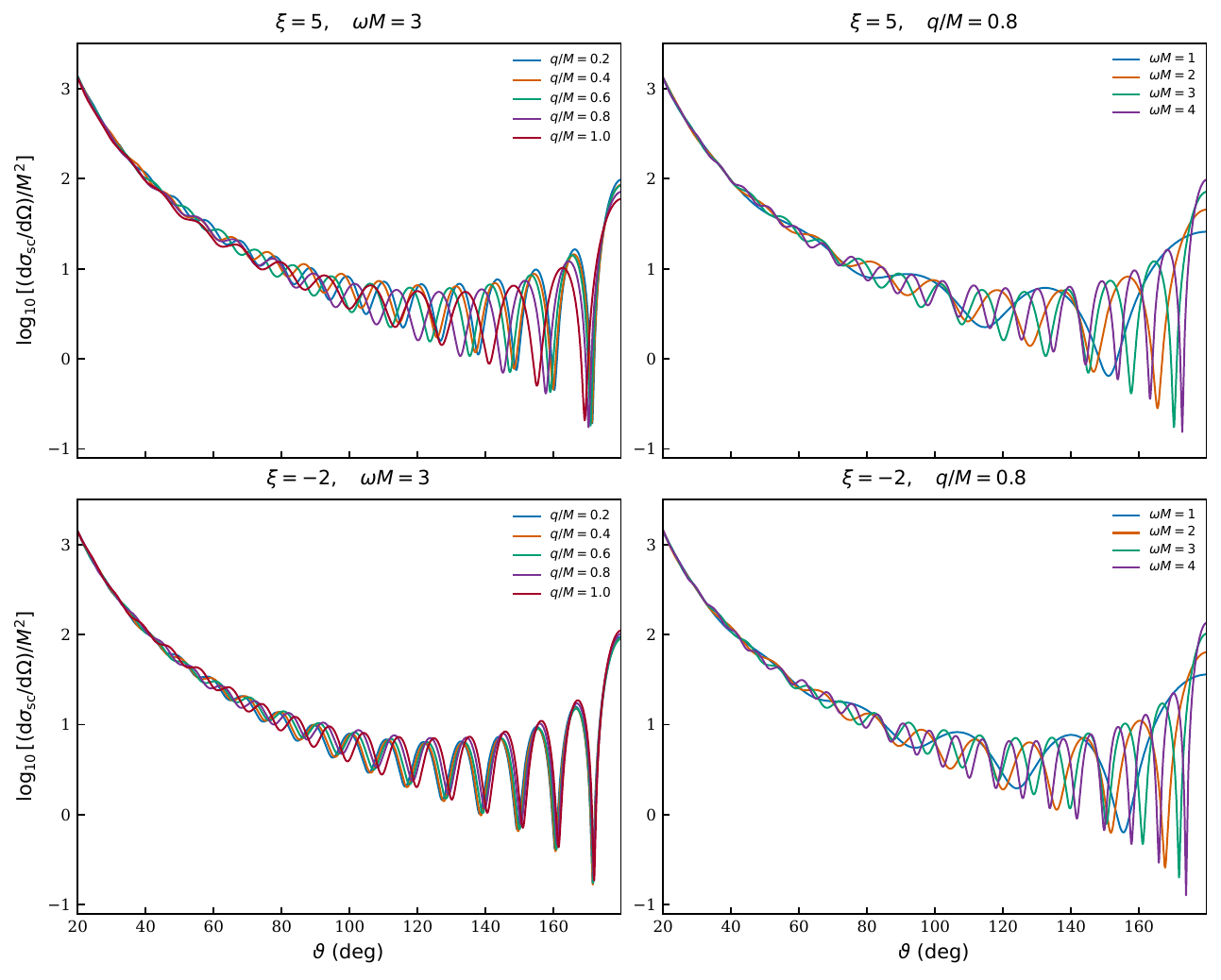}
  \captionof{figure}{Differential scattering $\log_{10}[M^{-2}\mathrm{d}\sigma_{\rm sc}/\mathrm{d}\Omega]$ as a function of $\vartheta$ at $M=1$. Upper row: $\xi=5$ and  lower row: $\xi=-2$. Left column: $\omega M=3$ with $q/M=0.2,0.4,0.6,0.8,1.0$. Right column: $q/M=0.8$ with $\omega M=1,2,3,4$. Colors identify the scanned parameter in each column. All four panels use the same axis ranges with $20^\circ\leq\vartheta\leq179.8^\circ$.}
  \label{fig:scattering-parameter-scans}
\end{figure}

Furthermore, Fig.~\ref{fig:scattering-parameter-scans} separates charge and frequency effects at fixed mass. In the left column, increasing $q/M$ widens the normal-branch backward fringes but slightly narrows the phantom-branch fringes. The change in fringe spacing is more pronounced in the normal branch. In the right column, increasing $\omega M$ narrows the fringes and raises the backward peak in both branches. The curves remain close at smaller angles, with larger differences appearing toward the backward region. The fringe spacing is consistent with the $\omega b_g$-dependence of the glory expression. Their common case $(q/M,\omega M)=(0.8,3)$ provides a shared reference spectrum for the two scans.

\FloatBarrier

\subsection{Normal BN--RN comparison at matched impact scales}
\begin{figure}[htbp]
  \centering
  \includegraphics[width=0.98\linewidth]{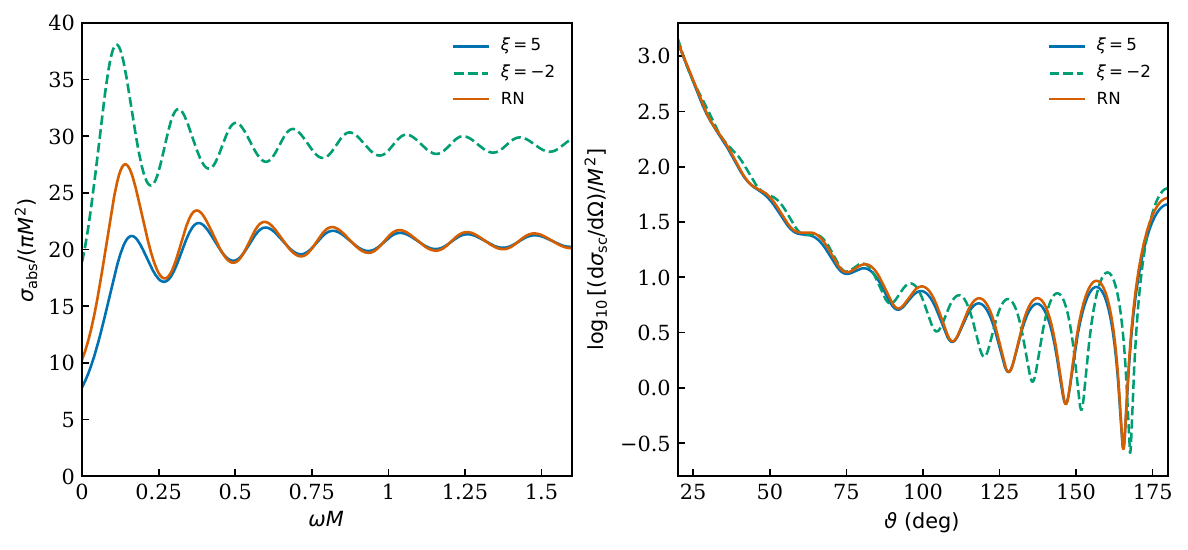}
  \caption{Normal BN ($\xi=5$, blue solid), phantom BN ($\xi=-2$, green dashed), and RN (orange solid) cross sections at $M=1$. Both BN branches use $q/M=0.8$. Left: total absorption $\sigma_{\rm abs}/(\pi M^2)$ for $0\leq\omega M\leq1.6$ and  RN $Q/M=0.7954$ matches the normal branch at $b_c/M\simeq4.6$. Right: $\log_{10}[M^{-2}\mathrm{d}\sigma_{\rm sc}/\mathrm{d}\Omega]$ at $\omega M=2$. The independent RN reference has $Q/M=0.8225$, matching the normal branch at $b_g/M\simeq4.7$. The phantom curves provide the same-mass, same-$q/M$ branch comparison but  are not matched to RN.}
  \label{fig:bn-rn-comparison}
\end{figure}

It is useful to compare the normal branch BN and RN cross sections at matched impact scales, as shown in Fig.~\ref{fig:bn-rn-comparison}. The RN metric is $F_{\rm RN}=B_{\rm RN}=1-2M/r+Q^2/r^2$, with outer horizon $r_+=M+\sqrt{M^2-Q^2}$. It is worth noting that matching $b_c$ aligns the geometric absorption limit and leading sinc oscillation scale. Matching $b_g$ instead aligns the leading glory fringe spacing at fixed frequency.

\FloatBarrier

Importantly, we observe that the normal BN and RN absorption spectra oscillate about a common geometric limit. The normal BN black hole has a lower zero-frequency limit because its horizon is smaller. The scattering spectra have similar backward fringe spacing but different amplitudes.

At the same mass and scalar charge, the phantom branch has a larger horizon area and capture scale, raising the low-frequency and geometric absorption levels. Its different glory scale changes the angular fringe spacing. For an RN black hole at $M=1$, one has $b_c/M\leq3\sqrt{3}\simeq5.1962$ and $b_g/M\leq5.3570$. However,  it is noted that  the phantom values $b_c/M=5.4070$ and $b_g/M=5.5695$ lie outside these ranges, preventing either matching at the same mass.

On the other hand, two matches require different RN charges at $q/M=0.8$ because $b_c$ and $b_g$ depend differently on the geometry. Matching either scale does not ensure agreement of the horizon areas, wave potentials, or glory prefactors. The remaining low- and finite-frequency differences therefore show that one geodesic scale does not determine the full scalar-wave response.

\FloatBarrier
\section{Conclusion}
\label{sec:conclusion}

We have studied massless scalar absorption and scattering in the normal and phantom BN spacetimes by employing null geodesics and partial waves. These black holes are curious because they  include  scalar hair charge $q$ instead of electric charge $Q$.
All comparisons fix the ADM mass $M=1$, with the horizon radius $r_{h,\xi}(M=1,q)$ determined separately for each branch with $\xi=5,-2$ and scalar charge $q$. The normal BN horizon radius $r_{h,5}$ is restricted to the $q$-range of $0\le q\le 18/17$ beyond the RN-range  of $0\le Q\le 1$, whereas there is not limit on $q$ for the phantom BN horizon $r_{h,-2}$. They have two single horizons, contrasting to double horizon for RN case. 

Over the sampled range of $q/M=0.2,0.4.0.6,0.8,1$, we observed that  increasing the charge reduces the horizon area and capture cross section and raises the peak of the $s(\ell=0)$-scalar mode potential in the normal branch.  However, we found that the phantom branch has the opposite trends to the normal branch.

The absorption spectra for both branches approach the horizon area at low frequency and oscillate about their geometric capture limit at high frequency. We observe that  photon-sphere normalization brings the spectra close together in both branches, with $\lambda_c/\Omega_c=1$ giving a common leading high-frequency reference.

On the other hand, for scattering, charge and branch shifted the interference pattern, while increasing frequency narrows the fringes.

Finally, the independently matched normal BN--RN comparisons aligned the leading capture or glory scale,  but they leave finite-frequency differences in the wave response.
This may  be due to sizable difference between horizon radii for larger scalar charge $q>0.5$. 

\section*{Acknowledgments}

 Y.S.Myung was supported by the National Research Foundation of Korea (NRF) grant funded by the Korea government(MSIT) (RS-2022-NR069013). 

\end{document}